\documentclass[11pt]{article}

\usepackage{pgfplots}
\usepackage{tikz}
\usepackage{tikzpeople}
\usepackage{circuitikz}
\usetikzlibrary{tikzmark, shapes, fit,backgrounds}
\usetikzlibrary{matrix, decorations.pathreplacing, angles, quotes}
\usetikzlibrary{patterns, calc, positioning}
\usetikzlibrary{arrows, arrows.meta}
\usetikzlibrary{shapes.multipart}
\usetikzlibrary{fit}
\usetikzlibrary{shapes.geometric, positioning}
\usetikzlibrary{decorations.pathmorphing}
\tikzstyle{block} = [rectangle, draw, 
    minimum width=4em, text centered, rounded corners, minimum height=1em]
\tikzstyle{line} = [draw, -latex]
\tikzset{meter/.append style={draw, inner sep=10, rectangle, font=\vphantom{A}, minimum width=30, scale=.7, path picture={\draw[black] ([shift={(.1,.3)}]path picture bounding box.south west) to[bend left=50] ([shift={(-.1,.3)}]path picture bounding box.south east);\draw[black,-{Latex[scale=.5]}] ([shift={(0,.1)}]path picture bounding box.south) -- ([shift={(.3,-.1)}]path picture bounding box.north);}}}
\tikzset{snake it/.style={decorate, decoration=snake}}

\usepackage[utf8]{inputenc}
\usepackage{amsmath}
\usepackage{amssymb}
\usepackage{amsthm}
\usepackage{color,soul}
\usepackage{tikz}
\usetikzlibrary{topaths,calc}
\usepackage{subfigure}
\usepackage{cite}

\usepackage{braket}

\newtheorem{theorem}{Theorem}

\newtheorem{remark}{Remark}

\newcommand\blfootnote[1]{%
  \begingroup
  \renewcommand\thefootnote{}\footnote{#1}%
  \addtocounter{footnote}{-1}%
  \endgroup
}

\title{Quantum Secret Sharing with Classical and Quantum Shares}
\author{Hua Sun}
\date{}

\begin{document}
\maketitle

\blfootnote{
Hua Sun (email: hua.sun@unt.edu) is with the Department of Electrical Engineering at the University of North Texas.}

\begin{abstract}
In quantum secret sharing, a quantum secret state is mapped to multiple shares such that shares from qualified sets can recover the secret state and shares from other forbidden sets reveal nothing about the secret state; we study the setting where there are both classical shares and quantum shares. We show that the quantum secret sharing problem with both classical and quantum shares is feasible if and only if any two qualified sets have some quantum share in common. Next, 
for threshold quantum secret sharing where there are $N_1$ classical shares, $N_2$ quantum shares and qualified sets consist of any $K_1$ (or more) classical shares and any $K_2 > N_2/2$ (or more) quantum shares, we show that to share $1$ qubit secret, each classical share needs to be at least $2$ bits and each quantum share needs to be at least $1$ qubit. Finally, we characterize the minimum share sizes for quantum secret sharing with at most $2$ classical shares and at most $2$ quantum shares. The converse proofs rely on quantum information inequalities and the achievable schemes use classical secret sharing, (encrypted) quantum secret sharing with only quantum shares, superdense coding, treating quantum digits as classical digits, and their various combinations.
\end{abstract}

\newpage
\allowdisplaybreaks
\section{Introduction}

How to encode a quantum file $Q_0$ into a classical random variable $Y$ and a quantum state $Q$ so that we may recover the original quantum file if both classical and quantum states are available and otherwise nothing is leaked? This is a canonical problem that we know the information theoretically optimal cost, i.e., to encode $1$ qubit of the quantum file securely, the classical variable $Y$ needs to be at least $2$ bits and the quantum state $Q$ needs to be at least $1$ qubit. There are two well known code constructions that may achieve this optimal cost. The first one is quantum teleportation \cite{Bennett_Brassard_Crepeau_Jozsa_Peres_Wootters}, 
where $Q$ is one half of a Bell pair and $Y$ is the outcome of the Bell measurement of $Q_0$ and the other half of the Bell pair. While quantum teleportation was not invented for the purpose of secure encoding a quantum file but instead communicating a quantum file with only classical communication and pre-shared entanglement \cite{Bennett_Brassard_Crepeau_Jozsa_Peres_Wootters}, it was realized long ago to give a secure encoding scheme as a by-product. The second one is quantum one-time pad \cite{Boykin_Roychowdhury}, also known as private quantum channel \cite{Ambainis_Mosca_Tapp_Wolf, Leung}, where $Y$ is set as two uniform bits, say $\alpha, \beta$ and $Q$ is obtained from passing $Q_0$ through quantum gates $X^\alpha Z^\beta$ (i.e., with probably $1/4$, doing nothing; with probably $1/4$, applying $X$ gate; with probability $1/4$, applying $Z$ gate; and with probability $1/4$, applying both $X$ and $Z$ gates); after such a mixture, $Q$ turns out to be maximally mixed such that security is guaranteed and quantum gates are reversible such that we may invert $Q$ to $Q_0$ given the knowledge of $\alpha, \beta$.

\subsection{Quantum Secret Sharing}
The starting point of this work is to understand the information theoretic optimality of the resource consumption of quantum teleportation and quantum one-time pad from a quantum information inequalities perspective (i.e., producing a proof that is based solely on Von Neumann entropy). To this end, we find not only an entropic proof (covered by Theorem \ref{thm:twin}), but also that the proof techniques generalize to the broader topic of quantum secret sharing \cite{Hillery_Buzek_Berthiaume, Cleve_Gottesman_Lo}, which is our main topic of study in this work and we discuss next.

Quantum secret sharing\footnote{In quantum secret sharing \cite{Hillery_Buzek_Berthiaume, Karlsson_Koashi_Imoto, Cleve_Gottesman_Lo}, the secret can be classical as well; however, in this work, we focus exclusively on quantum secrets.} refers to the problem where a quantum secret state $Q_0$ is encoded into $N_1$ classical shares $Y_1, \cdots, Y_{N_1}$ and $N_2$ quantum shares $Q_1, \cdots, Q_{N_2}$ such that if a set of shares are {\em qualified}, then the secret can be perfectly recovered and otherwise nothing is revealed about the secret. The secure encoding problem discussed at the beginning of this work is a quantum secret sharing problem where $N_1 = N_2 = 1$ and the only qualified set is both shares. Quantum secret sharing was introduced initially as the setting where all shares are quantum, i.e., $N_1 = 0$ \cite{Hillery_Buzek_Berthiaume, Karlsson_Koashi_Imoto} and it was later realized \cite{Nascimento_Mueller_Imai} that some quantum shares can be downgraded to classical shares without affecting the feasibility (resource consumption might change, e.g., for the secure encoding problem, when $Y$ can be quantum, then $1$ qubit, instead of $2$ bits, suffices \cite{Hillery_Buzek_Berthiaume}), which observation is interesting because classical shares are much simpler to handle in practice. Previous work in the domain of quantum secret sharing with both classical shares and quantum shares mainly start from the setting with only quantum shares and see which shares can be made classical through new code constructions \cite{Nascimento_Mueller_Imai, Singh_Srikanth, Fortescue_Gour}. In this work, an alternative perspective is taken where we assume that the number of classical shares and quantum shares (i.e., $N_1, N_2$) are fixed a priori and study the fundamental information theoretic limits (both code constructions and impossibility claims), i.e., to share $1$ qubit secret, what are the minimum share sizes for both classical and quantum shares (i.e., the share rate region problem).

We next give an overview of the main results obtained. First consider the question - when is quantum secret sharing feasible? A previously known sufficient condition (Theorem 1 in \cite{Nascimento_Mueller_Imai}) is shown to be necessary, thus fully establishing the feasibility condition, i.e., quantum secret sharing is feasible if and only if any two qualified sets of secret shares have at least one quantum share in common. This is reminiscence of the no-cloning property of quantum mechanics \cite{Wootters_Zurek, Dieks} and is similar to the feasibility condition of quantum secret sharing with only quantum shares \cite{Gottesman, Smith}. Our necessity converse proof is also similar, i.e., a minor generalization to include classical shares. Next we proceed to a previously studied threshold quantum secret sharing problem (called twin threshold \cite{Singh_Srikanth}) and show that a previously known code construction (refer to Section V.A in \cite{Singh_Srikanth}) is information theoretically optimal, i.e., minimizing both classical and quantum share sizes. In threshold quantum secret sharing, 
a qualified set 
consist of at least $K_1$ out of $N_1$ classical shares and at least $K_2$ out of $N_2$ quantum shares. An interesting combination of quantum one-time pad, threshold classical 
and quantum secret sharing 
achieves the same performance as that when $N_1= K_1 = N_2 = K_2 = 1$, i.e., for a $1$ qudit secret, each classical share has $2$ $q$-ary digits and each quantum share has $1$ qudit - a classical (quantum) $K_1$ ($K_2$) out of $N_1$ ($N_2$) secret sharing scheme stores the classical key $\alpha, \beta$ (encrypted quantum state $X^\alpha Z^\beta Q_0$) in classical (quantum) shares.
Our contribution is on the converse side, i.e., we show that the 
share sizes 
are optimal. 
Finally, we consider the settings with 
$N_1 \leq 2$ classical shares and $N_2 \leq 2$ quantum shares. It turns out that to settle the share rate region for all such cases, we need to combine classically encrypted quantum secret sharing with different forms of classical secret sharing, superdense coding, and treating qudits as classical digits regarding achievable schemes. Furthermore, the optimal share rate region consists of extreme points where one quantum share has (normalized) $1$ qudit while the other quantum share has $2$ qudits, necessitating a new converse bound. 


\section{Problem Statement}
Consider a quantum secret $Q_0$ that consists of $\lambda_0$ qudits and each qudit lies in a $q$-dimensional Hilbert space. $Q_0$ is maximally entangled with a reference system $R$ such that $R Q_0$ is a pure state. A quantum secret sharing scheme encodes $Q_0$ through a completely positive, trace-preserving (CPTP) mapping (and $R$ goes through an identity mapping) 
to a classical-quantum state 
\begin{eqnarray}
\rho_{Y_1 \cdots Y_{N_1} R Q_1 \cdots Q_{N_2}} &=& \sum_{y_1, \cdots, y_{N_1}} p_{Y_1\cdots Y_{N_1}}(y_1, \cdots, y_{N_1}) \ket{y_1\cdots y_{N_1}} \bra{y_1 \cdots y_{N_1}}_{Y_1 \cdots Y_{N_1}} \otimes \rho_{R Q_1 \cdots Q_{N_2}}^{(y_1 \cdots y_{N_1})} \notag \\
&\triangleq& \rho \label{state}
\end{eqnarray}
where $p_{Y_1\cdots Y_{N_1}}(y_1, \cdots, y_{N_1})$ denotes the probability of $(Y_1, \cdots, Y_{N_1}) = (y_1, \cdots,  y_{N_1})$, 
and each $Y_i, i \in \{1,\cdots, N_1\} \triangleq [N_1]$ consists of $\lambda^Y_{i}$ $q$-ary classical discrete random variables. 
$\rho_{RQ_1 \cdots Q_{N_2}}^{(y_1 \cdots y_{N_1})}$ is the density matrix of the quantum system $R Q_1 \cdots Q_{N_2}$ when $(Y_1, \cdots, Y_{N_1}) = (y_1, \cdots,  y_{N_1})$, and each $Q_j, j \in [N_2]$ consists of $\lambda^Q_j$ $q$-dimensional qudits. We refer to $Y_i$ as a classical share and $Q_j$ as a quantum share.

A quantum secret sharing scheme needs to satisfy two constraints. The first constraint is called the decoding constraint; it says that if some shares form a set $D$ called a {\em qualified set}, then the quantum secret (and its entanglement with the reference system) can be perfectly recovered. Without loss of generality, we assume there are no redundant shares, i.e., any share appears in at least one qualified set.
A qualified set is called a {\em minimal} qualified set if it is not a proper subset of another qualified set. The set of all minimal qualified sets is denoted as $\mathcal{D}$ and the set of all qualified sets is called the {\em access structure}, denoted as $\mathcal{A}$, and defined as 
\begin{eqnarray}
\mathcal{A} = \mathcal{D} \cup \bar{\mathcal{D}} 
\end{eqnarray}
where $\bar{\mathcal{D}}$ is the set of all share sets that include minimal qualified sets as proper subsets, i.e., $\forall D \subset \mathcal{P}$ such that there exists $\underline{D} \in \mathcal{D}$ and $\underline{D} \subsetneq D$, we have $D \in \bar{\mathcal{D}}$, where $\mathcal{P} = 2^{\{Y_1, \cdots, Y_{N_1}, Q_1, \cdots, Q_{N_2}\}}$ is the power set of all shares. The decoding constraint is expressed entropically as \cite{Schumacher_Nielsen, Imai_Muller_Nascimento_Tuyls_Winter}
\begin{eqnarray}
I(R; D)_\rho = 2 H(R)_\rho, ~\forall D \in \mathcal{A}. \label{dec}
\end{eqnarray}
All Von Neumann entropy and mutual information terms in this work are measured in $q$-ary units. 

The second constraint is called the security constraint; it says that for any shares that do not form a qualified set (such a share set is called a {\em forbidden set}), then absolutely no information is revealed about the quantum secret. The security constraint is expressed entropically\footnote{(\ref{sec}) indicates that $R$ and $S$ are in a product state so that the shares in a forbidden set have the same density matrix for any quantum secret state \cite{Imai_Muller_Nascimento_Tuyls_Winter}.} as \cite{Imai_Muller_Nascimento_Tuyls_Winter}
\begin{eqnarray}
I(R; S)_\rho = 0, ~\forall S \notin \mathcal{A}, S \in \mathcal{P}. \label{sec}
\end{eqnarray}

The storage performance of a quantum secret sharing scheme is measured by the {\em share rates}, defined as follows.
\begin{eqnarray}
&& R^Y_i = \frac{\lambda^Y_i}{\lambda_0}, i \in [N_1] \label{ry}\\
&& R^Q_j = \frac{\lambda^Q_j}{\lambda_0}, j \in [N_2] \label{rq}
\end{eqnarray}
where $R^Y_i $ ($R^Q_j $) characterizes the number of classical (quantum) digits contained in the classical (quantum) share for every qudit of the quantum secret.

A share rate {\em tuple} $(R^Y_1, \cdots, R^Y_{N_1}, R^Q_1, \cdots, R^Q_{N_2} )$ is said to be {\em achievable} if there exists a quantum secret sharing scheme such that the decoding constraint (\ref{dec}) and the security constraint (\ref{sec}) are satisfied, and the share rates are component-wise smaller than or equal to $(R^Y_1, \cdots, R^Y_{N_1}, R^Q_1, \cdots, R^Q_{N_2} )$. The closure of the set of all achievable share rate tuples for all $\lambda_0, q$ is called the optimal share rate {\em region}, denoted as $\mathcal{R}^*$.

\section{Results}
Denote the set of all quantum shares as $\mathcal{Q} = \{Q_1, \cdots, Q_{N_2}\}$ and the feasibility condition of quantum secret sharing is stated in the following theorem.

\begin{theorem}\label{thm:iff}
For quantum secret sharing with any access structure $\mathcal{A}$, 
\begin{eqnarray}
\mathcal{R}^* \neq \emptyset &\Leftrightarrow& \forall D_1, D_2 \in \mathcal{A}, D_1 \cap D_2 \cap \mathcal{Q} \neq \emptyset.  \label{eq:iff}
\end{eqnarray}
\end{theorem}

The proof of Theorem \ref{thm:iff} is presented in Section \ref{proof:iff}. When there is no classical shares, i.e., $N_1=0$, Theorem \ref{thm:iff} recovers the feasibility condition of quantum secret sharing with only quantum shares \cite{Gottesman, Smith}.
\vspace{0.1in}

Denote the set of classical shares as $\mathcal{Y} = \{Y_1, \cdots, Y_{N_1}\}$. Twin threshold quantum secret sharing refers to the following access structure,
\begin{eqnarray}
\mathcal{A} = \{D_Y \cup D_Q: D_Y \subset \mathcal{Y}, |D_Y| \geq K_1, D_Q \subset \mathcal{Q}, |D_Q| \geq K_2 > N_2/2 \} \label{twin}
\end{eqnarray}
and its optimal share rate region is characterized in the following theorem.

\begin{theorem}\label{thm:twin}
For twin threshold quantum secret sharing (refer to (\ref{twin})),
\begin{eqnarray}
\mathcal{R}^* = \Big\{(R^Y_1, \cdots, R^Y_{N_1}, R^Q_1, \cdots, R^Q_{N_2} ): R^Y_{i} \geq 2, R^Q_{j} \geq 1, i \in [N_1], j \in [N_2] \Big\} .\label{eq:twin}
\end{eqnarray}
\end{theorem}

The proof of Theorem \ref{thm:twin} is presented in Section \ref{proof:twin}. When $N_1 = K_1 = N_2 = K_2 = 1$, twin threshold quantum secret sharing reduces to the secure encoding problem discussed at the beginning of this work, thus Theorem \ref{thm:twin} covers the optimality of quantum teleportation and quantum one-time pad as a special case (with an entropic converse proof).

\vspace{0.1in}

Going beyond twin threshold access structure, it turns out that quantum secret sharing with $N_1 = 1$ classical share and $N_2 = 2$ quantum secret shares only has one non-redundant feasible access structure (specified through minimal qualified sets below), whose optimal share rate region is characterized in the following theorem.

\begin{theorem}\label{thm:21}
For quantum secret sharing with $N_1 = 1$ classical share, $N_2 = 2$ quantum shares and
\begin{eqnarray}
\mathcal{D} = \big\{ \{Y_1, Q_1\}, \{Q_1, Q_2\} \big\}, \label{21}
\end{eqnarray}
we have
\begin{eqnarray}
\mathcal{R}^* = \Big\{(R^Y_1, R^Q_1, R^Q_{2} ): R^Y_{1} \geq 2, R^Q_{1} \geq 1, R^Q_{2} \geq 1,  R^Q_{1} + R^Q_{2}\geq 3  \Big\} . \label{eq:21}
\end{eqnarray}
\end{theorem}

The proof of Theorem \ref{thm:21} is presented in Section \ref{proof:21}. The access structure (\ref{21}) is interesting because we have a solely quantum qualified set $\{Q_1, Q_2\}$ and a hybrid classical and quantum qualified set $\{Y_1, Q_1\}$, whose simultaneous presence demands a deeper interaction between classical and quantum resources while classically encrypted quantum secret sharing that works well for twin threshold access structure no longer suffices. New extreme points where $(R^Q_1, R^Q_2) = (1, 2)$ and $(R^Q_1, R^Q_2) = (2, 1)$ arise for which achievable schemes require new elements from superdense coding \cite{Bennett_Wiesner} and treating qudits as classical; converse bound $R^Q_{1} + R^Q_{2}\geq 3$ needs to use wisely the constraints from both solely quantum and hybrid classical-quantum qualified sets.
\vspace{0.1in}

Finally, we proceed to the cases with $N_1 = 2$ classical shares and $N_2 = 2$ quantum shares. While there are many feasible access structures that are not twin threshold, it turns out that interestingly, the optimal share region is the same for all of them and is presented in the following theorem.

\begin{theorem}\label{thm:22}
For quantum secret sharing with $N_1 = 2$ classical shares and $N_2 = 2$ quantum shares,
\begin{eqnarray}
\mathcal{R}^* = \Big\{(R^Y_1, R^Y_2, R^Q_1, R^Q_{2} ): R^Y_{1} \geq 2, R^Y_2 \geq 2, R^Q_{1} \geq 1, R^Q_{2} \geq 1,  R^Q_{1} + R^Q_{2}\geq 3  \Big\} \label{eq:22}
\end{eqnarray}
for the following access structures
\begin{eqnarray}
&1.& \mathcal{D} = \big\{  \{Y_1, Y_2, Q_1\}, \{Q_1, Q_2\} \big\} \notag \\
&2.& \mathcal{D} = \big\{  \{Y_1, Q_1\}, \{Y_2, Q_1, Q_2\} \big\}  \notag \\
&3.& \mathcal{D} = \big\{  \{Y_1, Y_2, Q_1\}, \{Y_2, Q_1, Q_2\} \big\} \notag \\
&4.& \mathcal{D} = \big\{  \{Y_1, Q_1\}, \{Y_2, Q_1\}, \{Q_1, Q_2\} \big\} \notag \\
&5.& \mathcal{D} = \big\{  \{Y_1, Y_2, Q_1\}, \{Y_1, Q_1, Q_2\} , \{Y_2, Q_1, Q_2\} \big\}. \label{22}
\end{eqnarray}
\end{theorem}

The proof of Theorem \ref{thm:22} is presented in Section \ref{proof:22}. (\ref{eq:22}) is similar to (\ref{eq:21}), so is the proof. Particularly, the five different access structures in (\ref{22}) require different twists of the achievable schemes of Theorem \ref{thm:21}.

\section{Proof of Theorem \ref{thm:iff}}\label{proof:iff}
The direction $\mathcal{R}^* \neq \emptyset \Leftarrow \forall D_1, D_2 \in \mathcal{A}, D_1 \cap D_2 \cap \mathcal{Q} \neq \emptyset$ has been proved in Theorem 1 of \cite{Nascimento_Mueller_Imai} and we only give an outline here while deferring the details to \cite{Nascimento_Mueller_Imai}. A quantum secret sharing scheme with non-zero rate is constructed as follows. Use any classical secret sharing scheme to share the uniform classical secret $(\alpha, \beta)$ for access structure $\mathcal{A}$ (where quantum shares are used as classical shares) which is known to exist \cite{Beimel, ito1989secret}. Consider only the quantum shares $\mathcal{Q}$ and use any quantum secret sharing scheme to share the maximally mixed quantum secret $X^\alpha Z^\beta Q_0$ for access structure $\mathcal{A} \cap \mathcal{Q}$ which is known to exist as any two qualified sets are intersecting, i.e., $D_1 \cap D_2 \cap \mathcal{Q} \neq \emptyset$ \cite{Gottesman, Smith}. As a result, from any qualified set $D \in \mathcal{A}$, we can recover $Q_0$ from $(\alpha, \beta)$ and $X^\alpha Z^\beta Q_0$; otherwise from any forbidden set $S \notin \mathcal{A}$, nothing about $(\alpha, \beta)$ and $X^\alpha Z^\beta Q_0$ (then $Q_0$) is revealed.

\vspace{0.1in}
The direction $\mathcal{R}^* \neq \emptyset \Rightarrow \forall D_1, D_2 \in \mathcal{A}, D_1 \cap D_2 \cap \mathcal{Q} \neq \emptyset$ is proved through its contrapositive statement, i.e., $\mathcal{R}^* = \emptyset \Leftarrow \exists D_1, D_2 \in \mathcal{A}, D_1 \cap D_2 \cap \mathcal{Q} = \emptyset$. Such a converse proof is presented next. For a set $\mathcal{J}$, $Q_{\mathcal{J}}$ denotes the set $\{Q_j : j \in \mathcal{J} \}$. Suppose $D_1 = \{Y_{\mathcal{I}_1}, Q_{\mathcal{J}_1}\}$, $D_2 = \{Y_{\mathcal{I}_2}, Q_{\mathcal{J}_2}\}$, then $\mathcal{J}_1 \cap \mathcal{J}_2 = \emptyset$. As $D_1$ is a qualified set, from the decoding constraint (\ref{dec}), we have
\begin{eqnarray}
2H(R)_\rho &\overset{(\ref{dec})}{=}& I(R; D_1)_\rho ~=~ I(R; Y_{\mathcal{I}_1}, Q_{\mathcal{J}_1})_\rho \\
&\leq& I(R; \mathcal{Y}, Q_{\mathcal{J}_1})_\rho \label{eq:sub1}
\end{eqnarray}
where (\ref{eq:sub1}) follows from the fact that quantum conditional mutual information is non-negative (equivalent to strong subadditivity/sub-modularity of quantum entropy functions) \cite{Nielsen_Chuang, Pippenger}, i.e., $ I(R; Y_{\mathcal{I}_1^c} \mid Y_{\mathcal{I}_1}, Q_{\mathcal{J}_1}) \geq 0$ (recall that $\mathcal{Y} = \{Y_1, \cdots, Y_{N_1}\}$ and $Y_{\mathcal{I}_1^c}$ is defined such that $\mathcal{Y} = Y_{\mathcal{I}_1} \cup Y_{\mathcal{I}_1^c}$). Symmetrically, for qualified set $D_2$, we have
\begin{eqnarray}
2H(R)_\rho &\leq& I(R; \mathcal{Y}, Q_{\mathcal{J}_2})_\rho. \label{eq:sub2}
\end{eqnarray}
Adding (\ref{eq:sub1}) and (\ref{eq:sub2}), we have
\begin{eqnarray}
4 H(R)_\rho &\leq& 2H(R)_\rho - \Big( H(R \mid \mathcal{Y}, Q_{\mathcal{J}_1})_\rho 
+ H(R \mid \mathcal{Y}, Q_{\mathcal{J}_2} )_\rho 
\Big) \\
&=& 2 H(R)_\rho - \sum_{y_1, \cdots, y_{N_1}}  p_{Y_1\cdots Y_{N_1}}(y_1, \cdots, y_{N_1}) \Big( H\big(R \mid Q_{\mathcal{J}_1}, Y_1 = y_1, \cdots, Y_{N_1} = y_{N_1} \big)_\rho \notag\\
&&~~+~ H\big(R \mid Q_{\mathcal{J}_2}, Y_1 = y_1, \cdots, Y_{N_1} = y_{N_1} \big)_\rho \Big)  \label{eq:cq1} \\
&=& 2 H(R)_\rho - \sum_{y_1, \cdots, y_{N_1}}  p_{Y_1\cdots Y_{N_1}}(y_1, \cdots, y_{N_1}) \Big( H\big(R \mid Q_{\mathcal{J}_1} \big)_{\rho_{RQ_1 \cdots Q_{N_2}}^{(y_1 \cdots y_{N_1})}} \notag\\
&&~~+~ H\big(R \mid Q_{\mathcal{J}_2} )_{\rho_{RQ_1 \cdots Q_{N_2}}^{(y_1 \cdots y_{N_1})}} \Big)  \label{eq:cq2} \\
&\leq& 2 H(R)_\rho \label{eq:weak1}\\
\Rightarrow~~ \lambda_0 &=& H(R)_\rho ~\leq~ 0 \label{eq:f1}
\end{eqnarray}
where (\ref{eq:cq1}) and (\ref{eq:cq2}) follow from the property of the entropy of a classical-quantum state (see e.g., Theorem 11.2.2 of \cite{Wilde}). 
(\ref{eq:weak1}) follows from weak monotonicity of quantum entropy \cite{Nielsen_Chuang, Pippenger, Wilde}. 
Finally, (\ref{eq:f1}) states that $\lambda_0 = 0$, i.e., the quantum secret must be null such that quantum secret sharing is not feasible. 

\section{Proof of Theorem \ref{thm:twin}}\label{proof:twin}
The achievability of the share rate tuple $(R_1^Y, \cdots, R_{N_1}^Y, R_1^Q, \cdots, R_{N_2}^Q) = (2, \cdots, 2, 1, \cdots, 1)$ has been proved in Section V.A of \cite{Singh_Srikanth} and we only give an outline here while deferring the details to \cite{Singh_Srikanth}. The proof is similar to that of Theorem \ref{thm:iff}, i.e., use classically encrypted quantum secret sharing and replace general classical/quantum secret sharing with the threshold version. Specifically, consider only the classical shares $\mathcal{Y}$ and use any $K_1$ out of $N_1$ classical threshold secret sharing scheme \cite{Shamir, Blakley, Beimel} to share the uniform $q$-ary classical secret $(\alpha, \beta)$ where each classical share has $2$ $q$-ary digits; consider only the quantum shares $\mathcal{Q}$ and use any $K_2 > N_2/2$ (otherwise the problem is infeasible according to Theorem \ref{thm:iff}) out of $N_2$ quantum threshold secret sharing scheme to share the maximally mixed $q$-dimensional quantum secret $X^\alpha Z^\beta Q_0$ where $q \geq 2K_2 -1$ is a prime power \cite{Cleve_Gottesman_Lo}, $Q_0$ has $1$ $q$-dimensional qudit and each quantum share has $1$ qudit. It is straightforward to verify that decoding and security constraints are satisfied.

\vspace{0.1in}
Next we prove the converse bounds $R_1^Y \geq 2$, $R_{1}^Q \geq 1$ (other $R_i^Y$ and $R_j^Y$ bounds follow similarly). First, we prove $R_1^Y \geq 2$. Consider qualified set $D = \{Y_1, Y_2, \cdots, Y_{K_1}, Q_1, \cdots, Q_{K_2} \}$ and forbidden set $S = \{Y_2, \cdots, Y_{K_1}, Q_1, \cdots, Q_{K_2} \}$.
\begin{eqnarray}
2H(R)_\rho &\overset{(\ref{dec})}{=}& I(R; D)_\rho ~=~ I(R; Y_1, S)_\rho \label{eq:t1}\\
&\overset{(\ref{sec})}{=}& I(R; Y_1 \mid S)_\rho ~=~ H(Y_1 \mid S)_\rho - H(Y_1 \mid R, S)_\rho \\
&\leq& H(Y_1)_\rho \label{eq:tc1}\\
&=& H(Y_1)  \label{eq:tc0}\\
&\leq& \lambda_1^Y \label{eq:tc2} \\
\Rightarrow ~~ 2\lambda_0 &=& 2H(R)_\rho ~\leq~ \lambda_1^Y \label{eq:tc3} \\
\Rightarrow ~~ R_1^Y &\overset{(\ref{ry})}{=}& \lambda_1^Y / \lambda_0 ~\geq~ 2 \label{eq:t2}
\end{eqnarray}
where (\ref{eq:tc1}) follows from the fact that quantum mutual information is non-negative, i.e., $I(Y_1; S)_\rho = H(Y_1)_\rho - H(Y_1 \mid S)_\rho \geq 0$ and the property that $H(Y_1 \mid R, S)_\rho \geq 0$ for the classical-quantum state $\rho$ where $Y_1$ belongs to the classical part (see Theorem 11.10 of \cite{Nielsen_Chuang} or Exercise 11.7.9 of \cite{Wilde}). (\ref{eq:tc0}) is due to the fact that $H(Y_1)_\rho$ (the quantum entropy of the reduced density matrix of $Y_1$ in the classical-quantum state $\rho$) is equal to $H(Y_1)$ (the Shannon entropy of classical random variable $Y_1$ whose distribution is the marginal of $p_{Y_1, \cdots, Y_{N_1}} (y_1, \cdots, y_{N_1})$). (\ref{eq:tc2}) is obtained because $Y_1$ contains $\lambda_1^Y$ $q$-ary classical variables and uniform distribution maximizes discrete Shannon entropy. (\ref{eq:tc3}) is obtained because the reference system $R$ is maximally mixed and has $\lambda_0$ qudits.

Second, we prove $R_1^Q \geq 1$. Consider qualified set $D = \{Y_1, \cdots, Y_{K_1}, Q_1, Q_2, \cdots, Q_{K_2} \}$ and forbidden set $S = \{Y_1, \cdots, Y_{K_1}, Q_2, \cdots, Q_{K_2} \}$.
\begin{eqnarray}
2H(R)_\rho &\overset{(\ref{dec})}{=}& I(R; D)_\rho ~=~ I(R; Q_1, S)_\rho \label{eq:t3} \\
&\overset{(\ref{sec})}{=}& I(R; Q_1 \mid S)_\rho = H(Q_1 \mid S)_\rho - H(Q_1 \mid R, S)_\rho \\
&\leq& H(Q_1)_\rho + H(Q_1)_\rho \label{eq:tc4} \\
&\leq& 2 \lambda_1^Q \label{eq:tc5}\\
\Rightarrow ~~ 2\lambda_0 &=& 2H(R)_\rho ~\leq~ 2\lambda_1^Q \\
\Rightarrow ~~ R_1^Q &\overset{(\ref{rq})}{=}& \lambda_1^Q / \lambda_0 ~\geq~ 1 \label{eq:t4}
\end{eqnarray}
where the second term of (\ref{eq:tc4}) follows from the Araki-Lieb inequality (see (11.73) of \cite{Nielsen_Chuang}) and (\ref{eq:tc5}) follows from the dimension bound for quantum entropy, i.e., $Q_1$ has $\lambda_1^Q$ qudits so that its entropy is upper bounded by $\lambda_1^Q$ (see Theorem 11.8 of \cite{Nielsen_Chuang}).

\begin{remark}
When we prove $R_1^Q \geq 1$ for quantum teleportation (i.e., $K_1 = N_1 = K_2 = N_2 = 1$), the above proof uses the assumption that $Y_1$ is independent of $R$ which is not necessary and a proof that does not rely on this assumption is as follows (while for the proof of $R_1^Y \geq 2$, the condition that $Q_1$ is independent of $R$ is required and also satisfied by teleportation as here $Q_1$ is pre-shared entanglement).
\begin{eqnarray}
2 H(R)_\rho &=& I(R; Y_1, Q_1)_\rho ~=~ H(R)_\rho - H(R \mid Y_1, Q_1)_\rho \\
&\leq& H(R)_\rho - H(R \mid Y_1, Q_1)_\rho + \underbrace{H(Q_1, R \mid Y_1)_\rho}_{\geq 0} \label{eq:te}\\
&=& H(R)_\rho  + H(Q_1 \mid Y_1)_\rho ~\leq~ H(R)_\rho  + H(Q_1)_\rho \\
&\leq& H(R)_\rho + \lambda_1^Q \\
\Rightarrow ~~~ \lambda_0 &=& H(R)_\rho ~\leq~ \lambda_1^Q \\
\Rightarrow ~~ R_1^Q &\overset{(\ref{rq})}{=}& \lambda_1^Q / \lambda_0 ~\geq~ 1
\end{eqnarray}
where the third term in (\ref{eq:te}) is non-negative because $Y_1RQ_1$ is a classical-quantum state.
\end{remark}

\section{Proof of Theorem \ref{thm:21}}\label{proof:21}
\subsection{Converse}
The proof of $R^Y_1 \geq 2$, $R_1^Q \geq 1$, $R_2^Q \geq 1$ is similar to that of Theorem \ref{thm:twin}. $R^Y_1 \geq 2$ can be proved by setting $D = \{Y_1, Q_1\}$, $S = \{Q_1\}$ and following (\ref{eq:t1})  to (\ref{eq:t2}). $R_1^Q \geq 1$ ($R_2^Q \geq 1$) can be proved by setting $D = \{Q_1, Q_2\}$, $S = \{Q_2\}$ ($S = \{Q_1\}$) and following (\ref{eq:t3}) to (\ref{eq:t4}).

The new converse bound $R_1^Q + R_2^Q \geq 3$ is proved next. Consider qualified sets $D_1 = \{Q_1, Q_2\}$, $D_2 = \{Y_1, Q_1\}$ and security set $S = \{Q_1\}$.
\begin{eqnarray}
4 H(R)_\rho &\overset{(\ref{dec})}{=}& I(R; D_1)_\rho + I(R; D_2)_\rho ~=~ I(R; Q_1, Q_2)_\rho + I(R; Y_1, Q_1)_\rho \label{eq:21c1}\\
&\overset{(\ref{sec})}{=}& I(R; Q_2 \mid Q_1)_\rho + I(R; Y_1, Q_1)_\rho \\
&=& H(Q_2 \mid Q_1)_\rho - H(Q_2 \mid Q_1, R)_\rho + H(R)_\rho - H(R \mid Y_1, Q_1)_\rho \\
&\leq& H(Q_2 \mid Q_1)_\rho - H(Q_2 \mid Y_1, Q_1, R)_\rho  - H(R \mid Y_1, Q_1)_\rho + H(R)_\rho \\
&=&  H(R)_\rho + H(Q_2 \mid Q_1)_\rho - H(R, Q_2 \mid Y_1, Q_1)_\rho \label{eq:21c}\\
&\leq&  H(R)_\rho + H(Q_2 \mid Q_1)_\rho - H(R, Q_2 \mid Y_1, Q_1)_\rho + \underbrace{ H(R, Q_1, Q_2 \mid Y_1)_\rho}_{\geq 0} \label{eq:21cc}\\
&=&  H(R)_\rho + H(Q_2 \mid Q_1)_\rho + H(Q_1 \mid Y_1)_\rho \\
&\leq& H(R)_\rho + H(Q_2 \mid Q_1)_\rho + H(Q_1)_\rho \\
&=& H(R)_\rho + H(Q_1, Q_2)_\rho \\
&\leq& H(R)_\rho + \lambda_1^Q + \lambda_2^Q\\
\Rightarrow ~~~~~~~~~ 3\lambda_0 &=& 3H(R)_\rho ~\leq~ \lambda_1^Q + \lambda_2^Q \\
\Rightarrow ~~ R_1^Q + R_2^Q &\overset{(\ref{rq})}{=}& (\lambda_1^Q + \lambda_2^Q) / \lambda_0 ~\geq~ 3 \label{eq:21c2}
\end{eqnarray} 
where (\ref{eq:21c}) follows from the definition of joint and conditional quantum entropy, i.e., the chain rule and in (\ref{eq:21cc}), the last term is non-negative because of the property of the entropy of a classical-quantum state (see e.g., Theorem 11.2.2 of \cite{Wilde}).

\subsection{Achievability}
The optimal share rate region $\mathcal{R}^*$ has two extreme points, $(R_1^Y, R_1^Q, R_2^Q) = (2, 1, 2)$ and $(R_1^Y, R_1^Q, R_2^Q)$ $= (2, 2, 1)$. For both cases, the achievability is based on the $K_1 = N_1 = K_2 = N_2 = 1$ quantum secret sharing problem and the quantum one-time pad code construction which we consider first. $\alpha, \beta$ are two uniform classical random variables over finite field $\mathbb{F}_q$ that are generated independently and thus in a product state with $R Q_0$.
\begin{eqnarray}
 \mathcal{D} = ~\{Y_1, Q_1\}, Y_1 = (\alpha, \beta), ~Q_1 = \mbox{apply}~X^\alpha Z^\beta~\mbox{to}~Q_0 \triangleq X^\alpha Z^\beta Q_0 \label{xz}
\end{eqnarray}
where the quantum gates $X, Z$ have well known generalizations to any finite field \cite{Wilde, Hayashi_Song, Yao_Jafar_ErasureMAC, aytekin2025quantum} ($\mathbb{F}_2$ suffices in this section, but $\mathbb{F}_3$ is needed for Theorem \ref{thm:22} in the next section).

Decoding constraint (\ref{dec}) holds because
\begin{eqnarray}
2H(R) \geq I(R; Y_1, Q_1)_\rho &\geq& I(R; \alpha, \beta, X^\alpha Z^\beta Q_0)_{\omega} \label{eq:21a1} \\
&=& I(R; X^\alpha Z^\beta Q_0 \mid \alpha, \beta)_{\omega} \label{eq:21a2} \\
&=& I(R; Q_0 \mid \alpha, \beta)_{\omega_{RQ_0}} ~=~ 2 H(R) \label{eq:21a3}
\end{eqnarray}
where (\ref{eq:21a1}) follows from quantum data processing inequality (see Theorem 11.9.4 of \cite{Wilde}) after measuring $Y_1$ to obtain $\alpha, \beta$ so that the classical-quantum state $\rho$ in (\ref{state}) becomes state $\omega$. (\ref{eq:21a2}) follows from the fact that $\alpha, \beta$ is generated independently of $R Q_0$. To obtain (\ref{eq:21a3}), we invert the quantum gates $X^\alpha Z^\beta$ and the state becomes the maximally entangled state $\omega_{RQ_0}$.

Security constraint (\ref{sec}) holds because $I(R; Y_1)_\rho = 0$ as $(\alpha, \beta)$ is generated independently of $R Q_0$ and $I(R; Q_1)_\rho = H(R)_\rho + H(Q_1)_\rho - H(R, Q_1)_\rho = 1 + 1 - 2 = 0$ because $Q_1$ and $R Q_1$ are both maximally mixed (the mixing of (\ref{xz}) is well known and for a proof, see Lemma 4 of \cite{Yao_Jafar_ErasureMAC}).

Next we are ready to prove the achievability of the two extreme points as follows, where the focus is on how to produce $\alpha, \beta$. Other non-extreme points in the optimal rate region can be achieved with time/space sharing.

\paragraph{Achievability of $(R_1^Y, R_1^Q, R_2^Q) = (2, 1, 2)$:}
Set
\begin{eqnarray}
Y_1 = (\alpha, \beta), ~Q_1 = X^\alpha Z^\beta Q_0, ~Q_2 = (\alpha, \beta)
\end{eqnarray}
so that $Q_2 = Y_1$, i.e., quantum share $Q_2$ is used as a classical share. Similar to above, from qualified set $\{Y_1, Q_1\}$ or $\{Q_1, Q_2\}$, we may recover the quantum secret and from other forbidden sets, nothing is leaked. The achieved rate values are as desired.

\paragraph{Achievability of $(R_1^Y, R_1^Q, R_2^Q) = (2, 2, 1)$:} Denote the superdense coding \cite{Bennett_Wiesner} of $(\alpha, \beta)$ as $\pi_1, \pi_2$, each of $1$ qudit, i.e., $\pi_0, \pi_1$ is a Bell pair and $\pi_2 = X^\alpha Z^\beta \pi_0$. From both ${\pi_1}, {\pi_2}$, we may recover $(\alpha, \beta)$; for any realizations of $(\alpha, \beta)$, each $\pi_1, \pi_2$ is maximally mixed \cite{Bennett_Wiesner, Nielsen_Chuang} thus each individual $\pi_1, \pi_2$ reveals nothing about $(\alpha, \beta)$ (similar to the security of quantum one-time pad considered above and superdense coding is well known to be intimately related to quantum one-time pad and quantum teleportation, see e.g., \cite{Boykin_Roychowdhury}). Set
\begin{eqnarray}
Y_1 = (\alpha, \beta), ~Q_1 = (X^\alpha Z^\beta Q_0, {\pi_1}), ~Q_2 = \pi_2
\end{eqnarray}
and from either $\{Y_1, Q_1\}$ or $\{Q_1, Q_2\}$, we may recover $Q_0$ and nothing is revealed about $Q_0$ for any forbidden set.

\section{Proof of Theorem \ref{thm:22}}\label{proof:22}
\subsection{Converse}
The converse proof is similar to that of Theorem \ref{thm:twin} and Theorem \ref{thm:21}. For $R_i^Y \geq 2, i = 1, 2$, we note that for any one of the five access structures in (\ref{22}), we may find a decoding set $D$ that includes $Y_i$ and removing $Y_i$ from $D$ gives a security set $S$; using this $D$ and $S$ in (\ref{eq:t1})  to (\ref{eq:t2}) gives us the $R_i^Y \geq 2$ bound. For $R_j^Q \geq 1, j = 1, 2$, we note that for all access structures in (\ref{22}), we may find a decoding set $D$ that includes $Q_j$ and removing $Q_j$ from $D$ gives a security set $S$; using this $D$ and $S$ in (\ref{eq:t3}) to (\ref{eq:t4}) proves $R_j^Q \geq 1$. For $R_1^Q + R_2^Q \geq 3$, consider each access structure in (\ref{22}) as follows. Set
\begin{eqnarray}
&1.& D_1 = \{Q_1, Q_2\}, ~D_2 = \{Y_1, Y_2, Q_1\}, ~S = \{Q_1\}\\
&2.& D_1 = \{Y_2, Q_1, Q_2\}, ~D_2 = \{Y_1, Y_2, Q_1\}, ~S = \{Y_2, Q_1\}\\
&3.& D_1 = \{Y_2, Q_1, Q_2\}, ~D_2 = \{Y_1, Y_2, Q_1\}, ~S = \{Y_2, Q_1\}\\
&4.& D_1 = \{Q_1, Q_2\}, ~D_2 = \{Y_1, Q_1\}, ~S = \{Q_1\}\\
&5.& D_1 = \{Y_2, Q_1, Q_2\}, ~D_2 = \{Y_1, Y_2, Q_1\}, ~S = \{Y_2, Q_1\}
\end{eqnarray}
and following (\ref{eq:21c1}) to (\ref{eq:21c2}) gives the desired $R_1^Q + R_2^Q \geq 3$ bound.

\subsection{Achievability}
We show that the extreme points $(R_1^Y, R_2^Y, R_1^Q, R_2^Q) = (2, 2, 1, 2) \triangleq \vec{R}_1$ and $(R_1^Y, R_2^Y, R_1^Q, R_2^Q) = (2, 2, 2, 1) \triangleq \vec{R}_2 $ are achievable for all access structures in (\ref{22}) (then all other points follow from space sharing). Similar to the achievability proof of Theorem \ref{thm:21}, generate $(\alpha, \beta)$ as two uniform classical random variables over $\mathbb{F}_3$. Generate independently another two uniform classical random variables $(\gamma_1, \gamma_2)$ over $\mathbb{F}_3$.

\begin{enumerate}
\item $\mathcal{D} = \big\{  \{Y_1, Y_2, Q_1\}, \{Q_1, Q_2\} \big\}$
\begin{eqnarray}
\vec{R}_1: && Y_1 = (\gamma_1, \gamma_2), Y_2 = (\alpha+\gamma_1, \beta+\gamma_2), Q_1 = X^\alpha Z^\beta, ~Q_2 = (\alpha, \beta) \\
\vec{R}_2: && Y_1 = (\gamma_1, \gamma_2), Y_2 = (\alpha+\gamma_1, \beta+\gamma_2), Q_1 = (X^\alpha Z^\beta, \pi_1), ~Q_2 = \pi_2
\end{eqnarray}
where $\pi_1, \pi_2$ is the superdense coding of $(\alpha, \beta)$.

\item $\mathcal{D} = \big\{  \{Y_1, Q_1\}, \{Y_2, Q_1, Q_2\} \big\}$
\begin{eqnarray}
\vec{R}_1: && Y_1 = (\alpha, \beta), Y_2 = (\alpha+\gamma_1, \beta+\gamma_2), Q_1 = X^\alpha Z^\beta, ~Q_2 = (\gamma_1, \gamma_2) \\
\vec{R}_2: && Y_1 = (\alpha, \beta), Y_2 = (\alpha+\gamma_1, \beta+\gamma_2), Q_1 = (X^\alpha Z^\beta, \pi_1), ~Q_2 = \pi_2
\end{eqnarray}
where $\pi_1, \pi_2$ is the superdense coding of $(\gamma_1, \gamma_2)$.

\item $\mathcal{D} = \big\{  \{Y_1, Y_2, Q_1\}, \{Y_2, Q_1, Q_2\} \big\}$
\begin{eqnarray}
\vec{R}_1: && Y_1 = (\gamma_1, \gamma_2), Y_2 = (\alpha+\gamma_1, \beta+\gamma_2), Q_1 = X^\alpha Z^\beta, ~Q_2 = (\gamma_1, \gamma_2) \\
\vec{R}_2: && Y_1 = (\gamma_1, \gamma_2), Y_2 = (\alpha+\gamma_1, \beta+\gamma_2), Q_1 = (X^\alpha Z^\beta, \pi_1), ~Q_2 = \pi_2
\end{eqnarray}
where $\pi_1, \pi_2$ is the superdense coding of $(\gamma_1, \gamma_2)$.

\item $\mathcal{D} = \big\{  \{Y_1, Q_1\}, \{Y_2, Q_1\}, \{Q_1, Q_2\} \big\}$
\begin{eqnarray}
\vec{R}_1: && Y_1 = (\alpha, \beta), Y_2 = (\alpha, \beta), Q_1 = X^\alpha Z^\beta, ~Q_2 = (\alpha, \beta) \\
\vec{R}_2: && Y_1 = (\alpha, \beta), Y_2 = (\alpha, \beta), Q_1 = (X^\alpha Z^\beta, \pi_1), ~Q_2 = \pi_2
\end{eqnarray}
where $\pi_1, \pi_2$ is the superdense coding of $(\alpha, \beta)$.

\item $\mathcal{D} = \big\{  \{Y_1, Y_2, Q_1\}, \{Y_1, Q_1, Q_2\} , \{Y_2, Q_1, Q_2\} \big\}$
\begin{eqnarray}
\vec{R}_1: && Y_1 = (\gamma_1, \gamma_2), Y_2 = (\alpha+\gamma_1, \beta+\gamma_2), Q_1 = X^\alpha Z^\beta, ~Q_2 = (\alpha+2\gamma_1, \beta+2\gamma_2) \\
\vec{R}_2: && Y_1 = (\gamma_1, \gamma_2), Y_2 = (\alpha+\gamma_1, \beta+\gamma_2), Q_1 = (X^\alpha Z^\beta, \pi_1), ~Q_2 = \pi_2
\end{eqnarray}
where $\pi_1, \pi_2$ is the superdense coding of $(\alpha+2\gamma_1, \beta+2\gamma_2)$.

\end{enumerate}

The proof that (\ref{dec}) and (\ref{sec}) hold is straightforward (specifically, note that $(\gamma_1, \gamma_2), (\alpha+\gamma_1, \beta+\gamma_2), (\alpha+2\gamma_1, \beta+2\gamma_2)$ is a $2$ out of $3$ threshold classical secret sharing of secret $(\alpha, \beta)$) and similar to that of Theorem \ref{thm:21}. The desired rate tuples are achieved.

\bibliography{Thesis}

\begin{thebibliography}{10}
\providecommand{\url}[1]{#1}
\csname url@samestyle\endcsname
\providecommand{\newblock}{\relax}
\providecommand{\bibinfo}[2]{#2}
\providecommand{\BIBentrySTDinterwordspacing}{\spaceskip=0pt\relax}
\providecommand{\BIBentryALTinterwordstretchfactor}{4}
\providecommand{\BIBentryALTinterwordspacing}{\spaceskip=\fontdimen2\font plus
\BIBentryALTinterwordstretchfactor\fontdimen3\font minus
  \fontdimen4\font\relax}
\providecommand{\BIBforeignlanguage}[2]{{%
\expandafter\ifx\csname l@#1\endcsname\relax
\typeout{** WARNING: IEEEtran.bst: No hyphenation pattern has been}%
\typeout{** loaded for the language `#1'. Using the pattern for}%
\typeout{** the default language instead.}%
\else
\language=\csname l@#1\endcsname
\fi
#2}}
\providecommand{\BIBdecl}{\relax}
\BIBdecl

\bibitem{Bennett_Brassard_Crepeau_Jozsa_Peres_Wootters}
C.~H. Bennett, G.~Brassard, C.~Cr{\'e}peau, R.~Jozsa, A.~Peres, and W.~K.
  Wootters, ``Teleporting an unknown quantum state via dual classical and
  {{Einstein-Podolsky-Rosen}} channels,'' \emph{Physical review letters},
  vol.~70, no.~13, p. 1895, 1993.

\bibitem{Boykin_Roychowdhury}
P.~O. Boykin and V.~Roychowdhury, ``Optimal encryption of quantum bits,''
  \emph{Physical review A}, vol.~67, no.~4, p. 042317, 2003.

\bibitem{Ambainis_Mosca_Tapp_Wolf}
A.~Ambainis, M.~Mosca, A.~Tapp, and R.~De~Wolf, ``Private quantum channels,''
  in \emph{Proceedings 41st Annual Symposium on Foundations of Computer
  Science}.\hskip 1em plus 0.5em minus 0.4em\relax IEEE, 2000, pp. 547--553.

\bibitem{Leung}
D.~W. Leung, ``{Quantum Vernam cipher},'' \emph{Quantum Information \&
  Computation}, vol.~2, no.~1, pp. 14--34, 2002.

\bibitem{Hillery_Buzek_Berthiaume}
M.~Hillery, V.~Bu{\v{z}}ek, and A.~Berthiaume, ``Quantum secret sharing,''
  \emph{Physical Review A}, vol.~59, no.~3, p. 1829, 1999.

\bibitem{Cleve_Gottesman_Lo}
R.~Cleve, D.~Gottesman, and H.-K. Lo, ``How to share a quantum secret,''
  \emph{Physical review letters}, vol.~83, no.~3, p. 648, 1999.

\bibitem{Karlsson_Koashi_Imoto}
A.~Karlsson, M.~Koashi, and N.~Imoto, ``Quantum entanglement for secret sharing
  and secret splitting,'' \emph{Physical Review A}, vol.~59, no.~1, p. 162,
  1999.

\bibitem{Nascimento_Mueller_Imai}
A.~C. Nascimento, J.~Mueller-Quade, and H.~Imai, ``Improving quantum
  secret-sharing schemes,'' \emph{Physical Review A}, vol.~64, no.~4, p.
  042311, 2001.

\bibitem{Singh_Srikanth}
S.~K. Singh and R.~Srikanth, ``Generalized quantum secret sharing,''
  \emph{Physical Review A—Atomic, Molecular, and Optical Physics}, vol.~71,
  no.~1, p. 012328, 2005.

\bibitem{Fortescue_Gour}
B.~Fortescue and G.~Gour, ``Reducing the quantum communication cost of quantum
  secret sharing,'' \emph{IEEE transactions on information theory}, vol.~58,
  no.~10, pp. 6659--6666, 2012.

\bibitem{Wootters_Zurek}
W.~K. Wootters and W.~H. Zurek, ``A single quantum cannot be cloned,''
  \emph{Nature}, vol. 299, no. 5886, pp. 802--803, 1982.

\bibitem{Dieks}
D.~Dieks, ``Communication by {{EPR}} devices,'' \emph{Physics Letters A},
  vol.~92, no.~6, pp. 271--272, 1982.

\bibitem{Gottesman}
D.~Gottesman, ``Theory of quantum secret sharing,'' \emph{Physical Review A},
  vol.~61, no.~4, p. 042311, 2000.

\bibitem{Smith}
A.~D. Smith, ``Quantum secret sharing for general access structures,''
  \emph{arXiv preprint quant-ph/0001087}, 2000.

\bibitem{Schumacher_Nielsen}
B.~Schumacher and M.~A. Nielsen, ``Quantum data processing and error
  correction,'' \emph{Physical Review A}, vol.~54, no.~4, p. 2629, 1996.

\bibitem{Imai_Muller_Nascimento_Tuyls_Winter}
H.~Imai, J.~M{\"u}ller-Quade, A.~C. Nascimento, P.~Tuyls, and A.~Winter, ``An
  information theoretical model for quantum secret sharing,'' \emph{Quantum
  Information \& Computation}, vol.~5, no.~1, pp. 69--80, 2005.

\bibitem{Bennett_Wiesner}
C.~H. Bennett and S.~J. Wiesner, ``Communication via one-and two-particle
  operators on {{Einstein-Podolsky-Rosen}} states,'' \emph{Physical review
  letters}, vol.~69, no.~20, p. 2881, 1992.

\bibitem{Beimel}
A.~Beimel, ``Secret-sharing schemes for general access structures: An
  introduction,'' \emph{Cryptology ePrint Archive}, 2025.

\bibitem{ito1989secret}
M.~Ito, A.~Saito, and T.~Nishizeki, ``Secret sharing scheme realizing general
  access structure,'' \emph{Electronics and Communications in Japan (Part III:
  Fundamental Electronic Science)}, vol.~72, no.~9, pp. 56--64, 1989.

\bibitem{Nielsen_Chuang}
M.~A. Nielsen and I.~L. Chuang, \emph{Quantum computation and quantum
  information}.\hskip 1em plus 0.5em minus 0.4em\relax Cambridge university
  press, 2010.

\bibitem{Pippenger}
N.~Pippenger, ``The inequalities of quantum information theory,'' \emph{IEEE
  Transactions on Information Theory}, vol.~49, no.~4, pp. 773--789, 2003.

\bibitem{Wilde}
M.~M. Wilde, \emph{Quantum information theory}.\hskip 1em plus 0.5em minus
  0.4em\relax Cambridge university press, 2013.

\bibitem{Shamir}
A.~Shamir, ``How to share a secret,'' \emph{Communications of the ACM},
  vol.~22, pp. 612--613, 1979.

\bibitem{Blakley}
G.~Blakley, ``Safeguarding cryptographic keys,'' in \emph{Proceedings of the
  1979 AFIPS National Computer Conference}, 1979, pp. 313--317.

\bibitem{Hayashi_Song}
M.~Hayashi and S.~Song, ``Unified approach to secret sharing and symmetric
  private information retrieval with colluding servers in quantum systems,''
  \emph{IEEE Transactions on Information Theory}, vol.~69, no.~10, pp.
  6537--6563, 2023.

\bibitem{Yao_Jafar_ErasureMAC}
Y.~Yao and S.~A. Jafar, ``Capacity of summation over a symmetric quantum
  erasure {{MAC}} with partially replicated inputs,'' \emph{IEEE Transactions
  on Information Theory}, 2025.

\bibitem{aytekin2025quantum}
A.~Aytekin, M.~Nomeir, S.~Vithana, and S.~Ulukus, ``{Quantum $X$-secure
  $E$-eavesdropped $T$-colluding symmetric private information retrieval},''
  \emph{IEEE Transactions on Information Theory}, 2025.

\end{thebibliography}

\end{document}